\documentclass[10pt, a4paper]{article}
\pdfoutput=1
\usepackage[utf8]{inputenc}
\usepackage[T1]{fontenc}
\usepackage{lmodern}
\usepackage{textcomp}
\usepackage[american]{babel}
\usepackage{microtype}
\usepackage{csquotes}

\usepackage{eurosym}
\usepackage{xspace}
\usepackage[nottoc]{tocbibind}
\usepackage{authblk}
\usepackage{totpages}

\usepackage{graphicx}
\usepackage{subcaption}
\usepackage[multiple]{footmisc}

\usepackage[usenames, dvipsnames, svgnames, table]{xcolor}

\usepackage[
	backend=bibtex8, citestyle=numeric-comp,
	sorting=none, sortcase=false, sortcites=true,
	giveninits=true, maxnames=99
]{biblatex}

\usepackage{amsmath, amsfonts, amssymb}
\usepackage{mathtools}
\usepackage{mathrsfs}
\usepackage{braket}
\usepackage{siunitx}

\usepackage[
	pdftex, breaklinks=true, linktocpage, colorlinks=true,
	urlcolor=RoyalBlue, linkcolor=RoyalBlue, citecolor=BrickRed
]{hyperref}
\usepackage[all]{hypcap}

\DeclareUnicodeCharacter{00A0}{~}
\DeclareUnicodeCharacter{202F}{~}

\newcommand{\email}[1]{\href{mailto:#1}{\nolinkurl{#1}}}
\newcommand{\emailfoot}[1]{\thanks{\email{#1}}}

\newcommand{\doi}[1]{\href{http://dx.doi.org/#1}{\nolinkurl{#1}}}
\newcommand{\urlhttp}[1]{\href{http://#1}{\nolinkurl{#1}}}

\newcounter{draftcommentcnt}
\NewDocumentCommand{\draftcomment}{s O{red} m}{%
	\def\margnote{\IfBooleanTF{#1}{\marginnote}{\marginpar}}%
	\stepcounter{draftcommentcnt}%
	\textcolor{#2}{#3}%
	\margnote{\textcolor{#2}{$\Leftarrow$ \arabic{draftcommentcnt}}}%
}

\graphicspath{{./images/}}
\numberwithin{equation}{section}

\usepackage[sort&compress, english]{cleveref}

\usepackage[heightrounded,
	top=3.5cm, bottom=3cm, left=3.5cm, right=3.5cm,
]{geometry}

\newcommand{\e}[0]{\mathrm{e}}

\newcommand{\N}[0]{\mathbb{N}}

\newcommand{\dd}[0]{\mathrm{d}}

\newcommand{\mc}[1]{{\mathcal{#1}}}

\newcommand{\mean}[1]{\langle #1 \rangle}

\newcommand{\abs}[1]{{|#1|}}

\input{arxiv.def}

\hypersetup{
	pdftitle={Renormalization in the neural network-quantum field theory correspondence},
	pdfauthor={Harold Erbin, Vincent Lahoche, Dine Ousmane Samary}
}

\title{Renormalization in the neural network-quantum field theory correspondence}

\author[1,2,3]{Harold Erbin\emailfoot{erbin@mit.edu}}

\author[1]{Vincent Lahoche\emailfoot{vincent.lahoche@cea.fr}}

\author[1,4]{Dine Ousmane Samary\emailfoot{dine.ousmanesamary@cea.fr}}

\affil[1]{%
	Université Paris-Saclay, CEA-LIST, Gif-sur-Yvette, F-91191, France
}

\affil[2]{%
	Center for Theoretical Physics, Massachusetts Institute of Technology
	\protect\\
	Cambridge, MA 02139, USA
}

\affil[3]{%
	NSF AI Institute for Artificial Intelligence and Fundamental Interactions
}

\affil[4]{%
	Faculté des Sciences et Techniques (ICMPA-UNESCO Chair)
	\protect\\
	Université d'Abomey-Calavi, 072 BP 50, Bénin
}

\begin{document}

\maketitle

\begin{abstract}
A statistical ensemble of neural networks can be described in terms of a quantum field theory (NN-QFT correspondence).
The infinite-width limit is mapped to a free field theory, while finite N corrections are mapped to interactions.
After reviewing the correspondence, we will describe how to implement renormalization in this context and discuss preliminary numerical results for translation-invariant kernels.
A major outcome is that changing the standard deviation of the neural network weight distribution corresponds to a renormalization flow in the space of networks.
\end{abstract}

\newpage

\hrule
\pdfbookmark[1]{\contentsname}{toc}
\tableofcontents
\bigskip
\hrule

\section{Introduction}
\label{sec:intro}

While neural networks (NN) perform extremely well on several tasks, they generally behave as black boxes which are hard to interpret~\cite{Weld:2018:ChallengeCraftingIntelligible,Zhang:2021:SurveyNeuralNetwork}.
This is a problem for applications where safety can be put in jeopardy~\cite{HighLevelExpertGrouponAI:2019:EthicsGuidelinesTrustworthy}, but also if concrete explanations are needed, as in sciences~\cite{Roscher:2020:ExplainableMachineLearning,Cranmer:2020:DiscoveringSymbolicModels,Raghu:2020:SurveyDeepLearning}.
Training is another concern because it is computationally expensive and has possible convergence issues.
Indeed, the loss function is typically non-convex such that it can be hard to find the global minimum~\cite{Choromanska:2015:LossSurfacesMultilayer,Li:2018:VisualizingLossLandscape}.
There is also no systematic hyperparameter tuning procedure and one has to rely on random scans, possibly improved with Bayesian and bandit methods~\cite{Bergstra:2012:RandomSearchHyperparameter,Snoek:2012:PracticalBayesianOptimization,Li:2018:HyperbandNovelBanditBased,Falkner:2018:BOHBRobustEfficient} which results in very high financial~\cite{Sharir:2020:CostTrainingNLP} and environmental costs~\cite{Strubell:2019:EnergyPolicyConsiderations,Lacoste:2019:QuantifyingCarbonEmissions,Henderson:2020:SystematicReportingEnergy}.
Finally, the question of knowing which functions can be expressed by a given NN remains open~\cite{Lu:2017:ExpressivePowerNeural,Raghu:2017:ExpressivePowerDeep}: while universal approximation theorems guarantee existence~\cite{Cybenko:1989:ApproximationSuperpositionsSigmoidal,Hornik:1989:MultilayerFeedforwardNetworks,Hornik:1991:ApproximationCapabilitiesMultilayer,Leshno:1993:MultilayerFeedforwardNetworks,Pinkus:1999:ApproximationTheoryMLP,Csaji:2001:ApproximationArtificialNeural}, finding the appropriate architecture for a new task often boils down to trials and errors.
Improving our theoretical understanding of NN is primordial for addressing these issues.

Physics provides a natural starting point for designing a theory of NN~\cite{Lin:2017:WhyDoesDeep,Zdeborova:2020:UnderstandingDeepLearning,Agliari:2020:MachineLearningStatistical}.
First, thanks to its effective descriptions, it is not necessary to know the fundamental theory.
Second, efficient representations of statistical models have been developed (path integrals, Feynman diagrams, statistical mechanics…).
Third, it allows characterizing the collective dynamics of degrees of freedom and organizing a phenomenon by scales.
Applications of physics to machine learning include statistical physics~\cite{Amit:1987:StatisticalMechanicsNeural,Gardner:1988:SpaceInteractionsNeural,Gardner:1988:OptimalStorageProperties,Krauth:1988:BasinsAttractionPerceptronlike,Mezard:1989:LearningFeedforwardLayered,Saitta:2011:PhaseTransitionsMachine,Choromanska:2015:LossSurfacesMultilayer,Bahri:2020:StatisticalMechanicsDeep}, renormalization~\cite{Beny:2013:DeepLearningRenormalization,Mehta:2014:ExactMappingVariational,Beny:2018:InferringRelevantFeatures,deMelloKoch:2020:DeepLearningRenormalization}, and QFT~\cite{Schoenholz:2017:CorrespondenceRandomNeural,Helias:2019:StatisticalFieldTheory,Halverson:2021:NeuralNetworksQuantum,Maiti:2021:SymmetryviaDualityInvariantNeural,Halverson:2021:BuildingQuantumField,Erbin:2022:NonperturbativeRenormalizationNeural,Grosvenor:2022:EdgeChaosQuantum}.

In this paper, we will review the neural network-quantum field theory (NN-QFT) correspondence developed in~\cite{Halverson:2021:NeuralNetworksQuantum,Erbin:2022:NonperturbativeRenormalizationNeural} since it provides concrete and testable tools to improve our analytical understanding of neural network building and training.
This correspondence states that, for a very general class of architectures, it is possible to associate a quantum field theory (QFT) with a statistical ensemble of NN.
We focus on a fully connected NN with a single hidden layer and setup non-perturbative renormalization group equations (valid for any finite width).
The main result is that varying the standard deviation of the weight distribution induces a renormalization group (RG) flow in the space of NN.
Code is available at: \url{https://github.com/melsophos/nnqft}.

\section{NN-QFT correspondence}
\label{sec:nnqft}

Take a fully connected neural network $f_{\theta,N}:
\mathbb{R}^{d_{\text{in}}} \rightarrow \mathbb{R}^{d_{\text{out}}}$ with one hidden layer of width $N$:
\begin{equation}
    f_{\theta,N}(x)
        = W_1\Big( g(W_0 x + b_0) \Big) + b_1,
\end{equation}
where $g$ is the non-linear activation function, and the parameters $\theta = (W_0, b_0, W_1, b_1)$ (weights and biases) have Gaussian distributions:
\begin{equation}
W_0 \sim \mc N(0, \sigma_W^2 / d_{\text{in}}),\quad
W_1 \sim \mc N(0, \sigma_W^2 / N),\quad
b_0, b_1 \sim \mc N(0, \sigma_b^2)\,.
\end{equation}
Consider next a statistical ensemble of neural networks, such that a given neural network is sampled from the distribution in parameter space: $f_{\theta, N} \sim P[\theta]$.
Then, there is a dual description in terms of another distribution in function space, which is induced by the parameter distribution plus the architecture: $f_{\theta, N} \sim p[f]$~\cite{Halverson:2021:NeuralNetworksQuantum}.
Changing the parameter distribution by training corresponds to flowing in the function space.

In the large $N$ limit (infinite width), the function distribution becomes a Gaussian process with kernel $K$ (as a consequence of the central limit theorem)~\cite{Neal:1996:BayesianLearningNeural}:
\begin{equation}
f \sim \mc N(0, K)\,.
\end{equation}
This statement generalizes to most architecture and training~\cite{Yang:2021:TensorProgramsWide}.
We denote as $S_0[f]$ the (Gaussian) log-probability:
\begin{equation}
    S_0[f]
        := \frac{1}{2} \int \dd^{d_{\text{in}}} x \dd^{d_{\text{in}}} x' \,
            f(x) \Xi(x, x') f(x'),
    \qquad
    \Xi := K^{-1}\,,
\end{equation}
and as
\begin{equation}
    G_0^{(n)}(x_1, \ldots, x_n)
        := \mathbb{E}_0 [ f(x_1) \cdots f(x_n)]
        \equiv \int \dd f \, \e^{- S_0[f]} \, f(x_1) \cdots f(x_n)
\end{equation}
the \emph{Gaussian expectation value} (GEV) for a product of $n$ fields $f(x_i)$.
The measure $\dd f$ is suitably normalized such that $\int \dd f \, \exp (- S_0[f]) = 1$.
In physics, this setting corresponds to a free QFT, $K$ to the free propagator and $G_0^{(n)}$ to the free $n$-point correlation functions (also called Green functions).
At finite $N$, the distribution is not a Gaussian process, and we denote as
\begin{equation}
    \Delta G^{(n)}:= G^{(n)} - G_0^{(n)}
\end{equation}
the difference between the \emph{full expectation value} (FEV)
\begin{equation}
    G^{(n)}(x_1, \ldots, x_n)
        := \mathbb{E}[ f(x_1) \cdots f(x_n)]
        \equiv \int \dd f \, \e^{- S[f]} \, f(x_1) \cdots f(x_n)
\end{equation}
and the GEV. The main message of the NN-QFT correspondence is that even at finite $N$, the log-probability $S[f]$ can be designed with non-Gaussian contributions to reproduce the FEVs with arbitrary precision up to the numerical uncertainties in the simulations. We denote as $S_{\text{int}}[f]$ the non-Gaussian contributions in $S[f]$:
\begin{equation}
    S[f] = S_0'[f] + S_{\text{int}}[f],
\end{equation}
where $S_0'[f] \neq S_0[f]$ is some new Gaussian action.
Indeed, the 2-point FEV $G^{(2)}(x, y)$ is $N$-independent and fixed by the NN, such that the Gaussian part must be different and such that:
\begin{equation}
    G^{(2)}(x, y)
        = G_0^{(2)}(x, y)
        \equiv K(x, y)\,.
\end{equation}

A complete dictionary between NN and QFT is given by Table~\ref{tab:dictionary}.
This formulation is promising because correlation functions between outputs give a measure of learning; e.g., the $1$-point function $\mathbb{E}[f(x)]$ corresponds to the average prediction for input $x$ (which is related to the idea of symmetry breaking in QFT~\cite{Maiti:2021:SymmetryviaDualityInvariantNeural}).
Hence, having a QFT may allow performing (semi-)analytic predictions in advance of the outcome of the learning process.

Kernels in data-space are typically bi-local~\cite{Halverson:2021:NeuralNetworksQuantum} such that one can expect non-local interactions.
Moreover, it is not clear what are the symmetries of the inputs and outputs (in the QFT sense) for general data. With these observations, we follow an approach which can be called NN phenomenology: 1) make assumptions dictated by numerical evidence, 2) write a QFT model to match observations, 3) use the model to check theoretical facts.

 \begin{table}[ht]
     \centering
     \begin{tabular}{c|c|c}

            & QFT
            & NN / GP
        \\
        \hline
        $x$ &
            spacetime points &
            data-space inputs
        \\
        $p$ &
            momentum space &
            dual data-space
        \\
        $f$ &
            field &
            neural network
        \\
        $K(x, y)$ &
            propagator &
            Gaussian kernel
        \\
        $S$ &
            action &
            negative log-probability
        \\
        $S_0$ &
            free action &
            Gaussian log probability
        \\
        $S_{\text{int}}$ &
            interactions &
            non-Gaussian corrections
        \\
        $\mean{\cdot} \equiv \mathbb E[\cdot]$ &
            expectation value, Green function &
            correlation function
    \end{tabular}
     \caption{NN-QFT dictionary.}
     \label{tab:dictionary}
 \end{table}

\section{Constructing the QFT}
\label{sec:results}

The expectation values $G^{(n)}$ can be computed analytically using QFT tools (“theory”) or computed from a statistical ensemble of neural networks (“measurements”).
Hence, we can make an ansatz for $S_{\text{int}}[f]$ and match the parameters by computing enough correlation functions.
The choice of this ansatz especially regarding symmetries and the way the fields are coupled depend on the Gaussian kernel $K$.
In this paper, we set $d_{\text{in}} = d_{\text{out}} = 1$ and focus on a translation-invariant activation function:
\begin{equation}
    g(W_0 x + b_0)
        = \frac{\exp(W_0 x + b_0)}{\sqrt{\exp \left[ 2 \Big( \sigma_b^2 + \frac{\sigma_W^2}{d_{\text{in}}} \, x^2 \Big) \right]}}
\end{equation}
such that the Gaussian kernel is~\cite{Halverson:2021:NeuralNetworksQuantum}:
\begin{equation}
    \label{eq:kernel}
    K(x, y)
        := \sigma_b^2 + K_W(x, y),
    \quad
    K_W(x, y)
        = \sigma_W^2 \, \e^{- \frac{\sigma_W^2}{2 d_{\text{in}}} \, \abs{x - y}^2}.
\end{equation}

\begin{figure*}
    \centering

    \begin{subfigure}[b]{0.45\linewidth}
        \centering
        \includegraphics[width=0.9\linewidth]{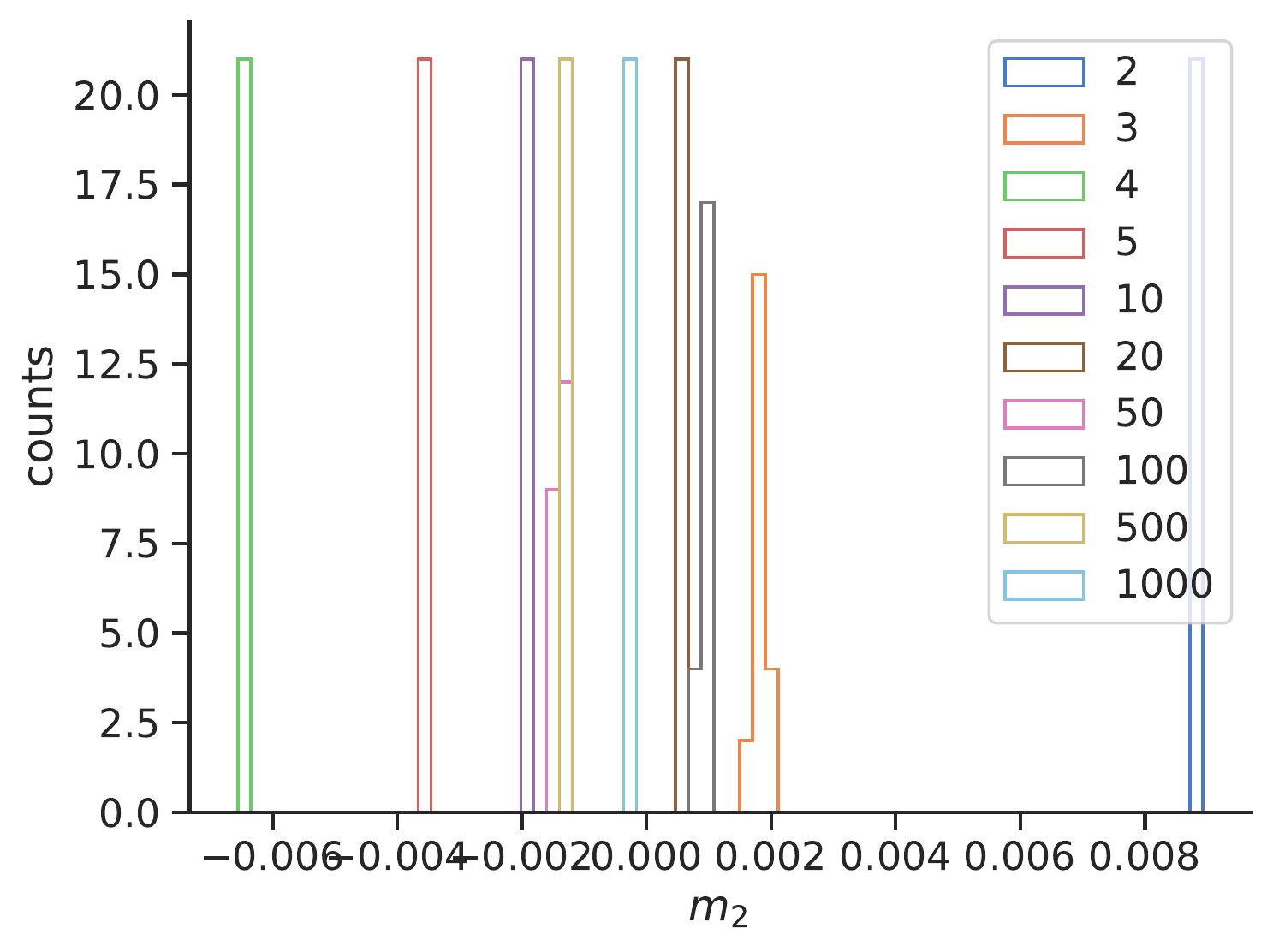}
        \caption{$m_2$.}
    \end{subfigure}
    \hfill
    \begin{subfigure}[b]{0.45\linewidth}
        \includegraphics[width=0.9\linewidth]{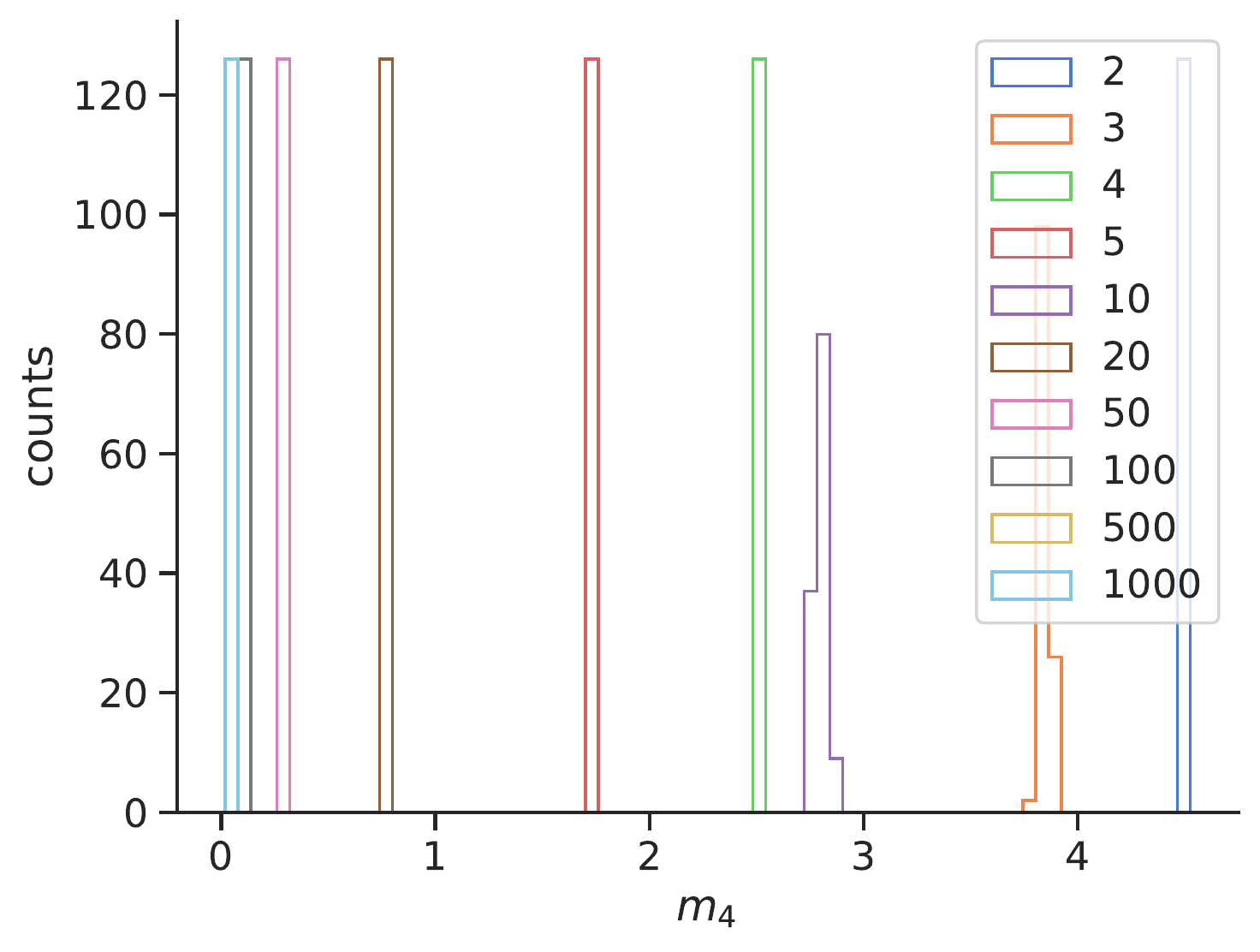}
        \caption{$m_4$.}
    \end{subfigure}

    \caption{
        Normalized deviations with respect to the free theory.
        For $m_2$, values centered around $0$ and independent of $N$.
        For $m_4$, the values decrease as $N$ increases.
        }
    \label{fig:hist-mn}
\end{figure*}

In order to compute the “experimental” Green functions for a given $N$, we create $n_{\text{bags}}$ distinct statistical ensembles of $n_{\text{nets}}$ networks each~\cite{Halverson:2021:NeuralNetworksQuantum,Erbin:2022:NonperturbativeRenormalizationNeural}, and compute $\bar G^{(n)}_{\text{exp}}$ as the average of the (empirical) FEV:
\begin{equation}
G^{(n)}_{\text{exp}} = \frac{1}{n_{\text{nets}}} \sum\limits_{\alpha=1}^{n_{\text{nets}}} f_\alpha(x_1) \cdots f_\alpha(x_n)\,,
\end{equation}
computed in a given bag. We furthermore define
\begin{equation}
\Delta G^{(n)}_{\text{exp}} := \bar G^{(n)}_{\text{exp}} - G_0^{(n)}\,,
\end{equation}
and the normalized deviation $m_n := \Delta G^{(n)}_{\text{exp}}/G_0^{(n)}$.
For the numerical investigations, we consider the points $x^{(1)}, \ldots, x^{(6)} \in \{ -0.01, -0.006, -0.002, 0.002, 0.006, 0.01 \}$ and evaluate the Green functions for all inequivalent combinations.
Moreover, all numerical tests are performed with $\sigma_b = 1, \sigma_W = 1, n_{\text{bags}} = 20, n_{\text{nets}} = 30000$, and $N \in \{ 2, 3, 4, 5, 10, 20, 50, 100, 500, 1000 \}$.
Computations ran during one week on the internal cluster of one of our institute.
Empirically, we find that $m_2 \approx 0$ (the second momentum is almost independent of $N$) and $m_{2n} = O\left( \frac{1}{N} \right)$ for $n > 1$, the last result meaning that the empirical $2n$-cumulant of the distribution $G_{c,\text{exp}}^{(2n)}$ must be of order $1/{N^{n-1}}$.
The histogram of values for $m_2$ and $m_4$ are given in Figure~\ref{fig:hist-mn}.

The translation invariance of the Gaussian kernel is reminiscent of standard QFTs, where $S_{\text{int}}$ can be expanded in powers of $f$ coupled at the same point, namely:
\begin{equation}
    S_{\text{int}}[f]
        := \sum_{n = 2}^{n_0} \frac{\bar{u}_n}{(2n)!} \int \dd^{d_{\text{in}}} x \, f(x)^{2n},
\end{equation}
for some $n_0 \in \N$.
We can check the validity of this ansatz experimentally.
Indeed, in that expression, $\bar{u}_n$ is nothing but the magnitude of the lowest order deviation from the GEV, and is called \textit{bare coupling}.
It is different from the effective coupling ${u}_n$ which is measured by the simulations and which includes quantum corrections; in perturbation theory (assuming $\bar{u}_n$ small enough), and $u_n = \bar{u}_n + \mathcal{O}(\bar{u}_n^2)$ (schematically). At higher order, this deviation receives many contributions which can be formally resumed.
For the lowest order, the full (normalized) deviation from the GEV $u_4(x_1, x_2, x_3, x_4)$ reads:
\begin{equation}
    \begin{gathered}
    u_{4}(x_1, x_2, x_3, x_4)
        = - \frac{\Delta G^{(4)}_{\text{exp}}(x_1, x_2, x_3, x_4)}{\int \dd^{d_{\text{in}}} x \, K_W(x, x_1) K_W(x, x_2) K_W(x, x_3) K_W(x, x_4)}.
    \end{gathered}
\end{equation}
Empirically, focusing on the truncation $n_0 = 3$, we find that $u_4$ is negative but almost constant and $u_6$ remains small but positive as required for stability.
Results for different $\sigma_W$ and $N$ are given below in Figure~\ref{fig:flow-u4-sw}.

\section{Renormalization group}
\label{sec:renorm}

In the previous section, we considered an \emph{effective field theory} able to reproduce FEV corresponding to a NN ensemble.
The RG is a set of techniques allowing to understand the dependency of the effective theory on a typical observation scale.
The machine precision provides an example of such an observation scale, and we could consider the dependency of the parameters defining the QFT regarding the machine precision.
In this paper, we consider another kind of scaling, induced by the NN itself and called \emph{active RG}.
The motivation stems from the observation that the propagator \eqref{eq:kernel} in momentum space looks like the usual Gaussian kernel in QFT at low momentum:
\begin{equation}
    \label{eq:kernel-p}
    K_W(p)
        = (\sigma_W^2)^{1 - \frac{d_{\text{in}}}{2}}
            \left( \frac{d_{\text{in}}}{2\pi} \right)^{\frac{d_{\text{in}}}{2}}
            \exp \left[ -\frac{d_{\text{in}}}{2\sigma_W^2} \, p^2 \right]
        \approx \frac{Z_0^{-1}}{\Lambda^2 + p^2 + O(p^2)}.
\end{equation}
In the QFT terminology, $\Lambda$ defines the \emph{mass scale}, and the large momenta $p^2 \gg \Lambda^2$ are exponentially suppressed, blinding the physics beyond scale $p^2\sim \Lambda^2$.
Hence, in the active RG, $\Lambda$ (or equivalently the standard deviation $\sigma_W$) define the typical momentum scale.

\begin{figure*}
    \centering

    \begin{subfigure}[b]{0.45\linewidth}
        \centering
        \includegraphics[width=0.9\linewidth]{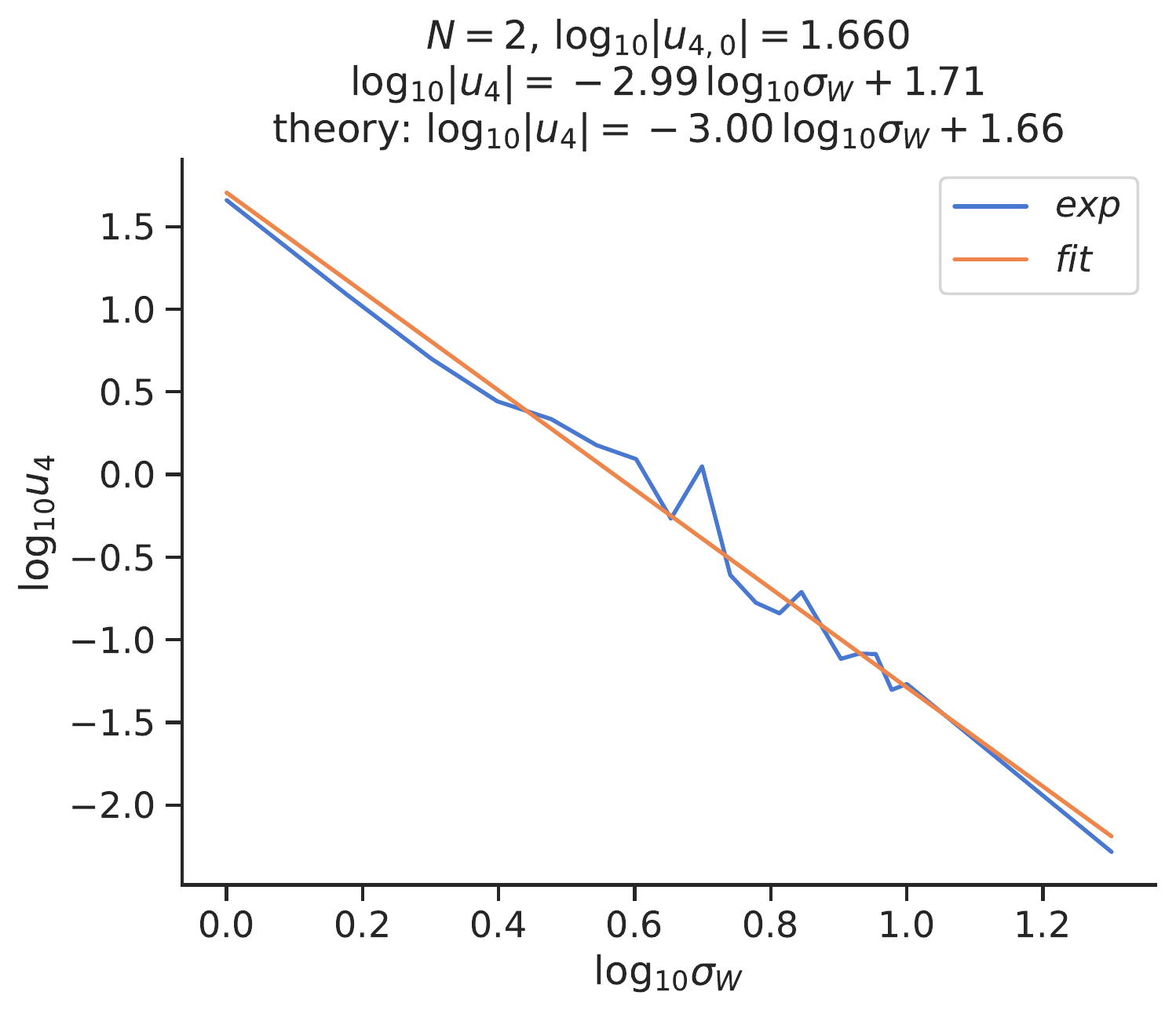}
        \caption{$N = 2$.}
    \end{subfigure}
    \hfill
    \begin{subfigure}[b]{0.45\linewidth}
        \includegraphics[width=0.9\linewidth]{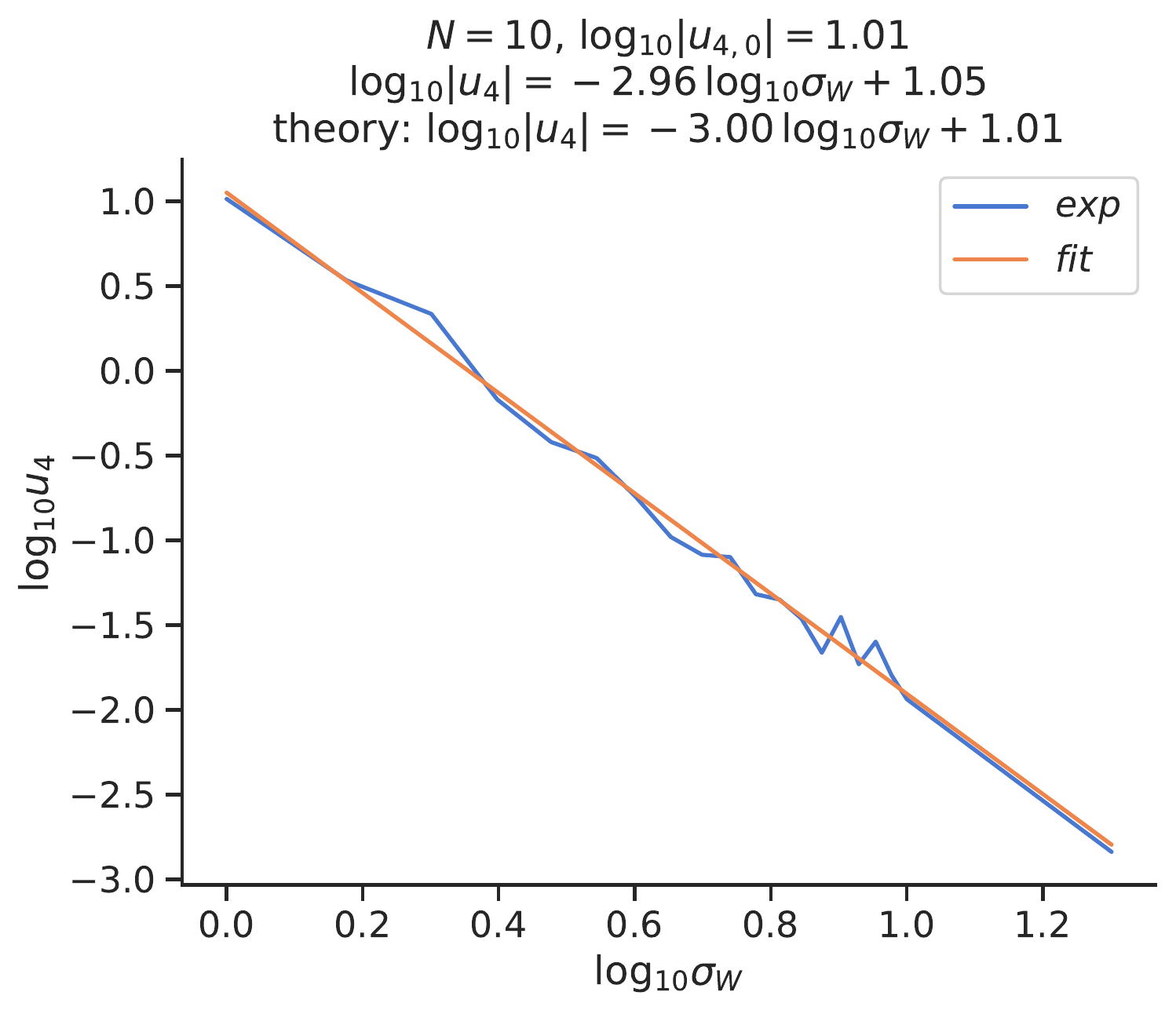}
        \caption{$N = 10$.}
    \end{subfigure}

    \begin{subfigure}[b]{0.45\linewidth}
        \centering
        \includegraphics[width=0.9\linewidth]{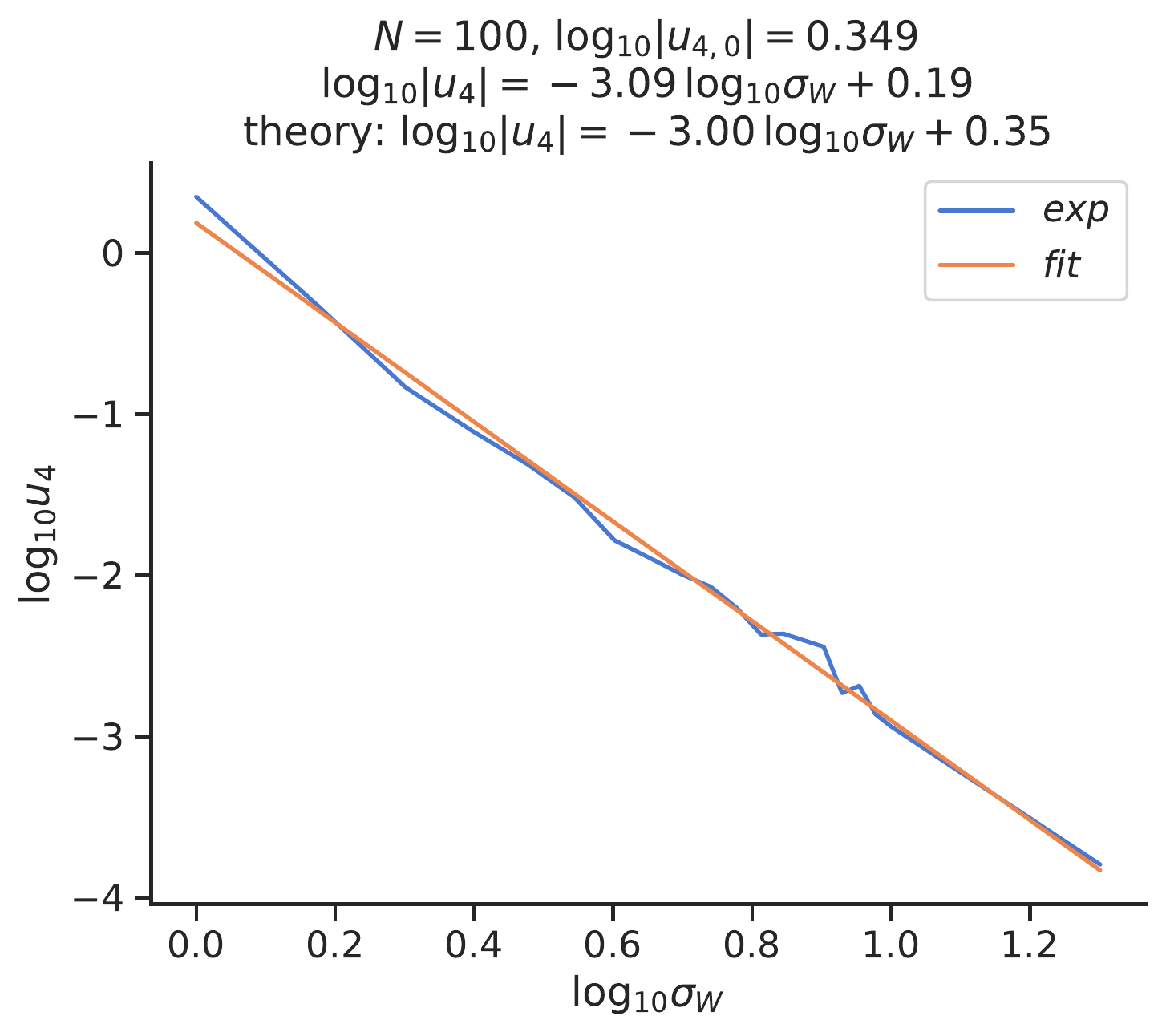}
        \caption{$N = 100$.}
    \end{subfigure}
    \hfill
    \begin{subfigure}[b]{0.45\linewidth}
        \includegraphics[width=0.9\linewidth]{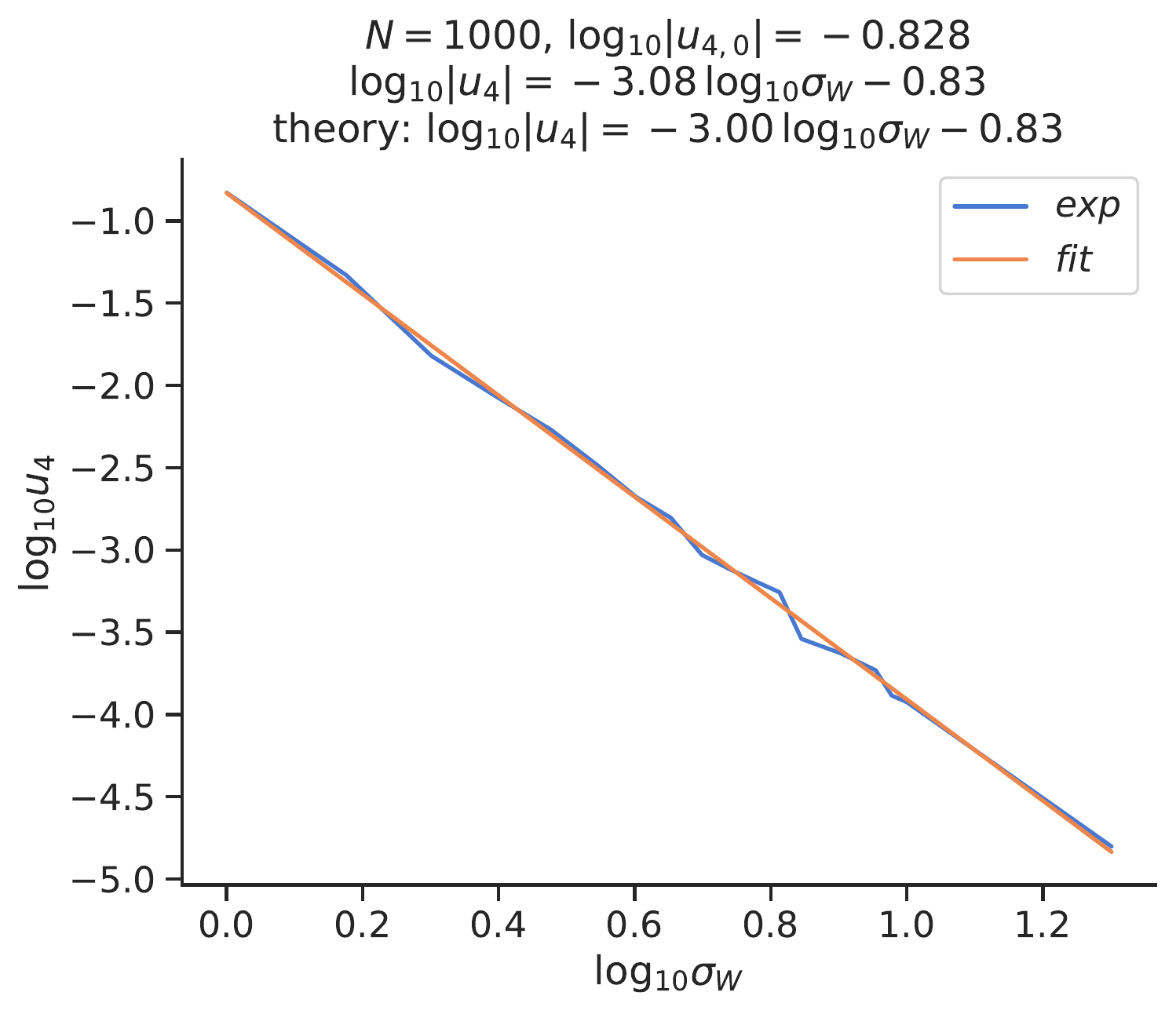}
        \caption{$N = 1000$.}
    \end{subfigure}

    \caption{
        Dependence of $u_4$ in terms of $\sigma_W$, computed numerically and with the flow equation \eqref{eq:flow-u4}.
        Parameters: $\sigma_b = 0, \sigma_W \in \{ 1.0, 1.5, \ldots, 10, 20 \}, n_{\text{bags}} = 30, n_{\text{nets}} = 30000$.
        }
    \label{fig:flow-u4-sw}
\end{figure*}

Having defined the notion of scale, we are aiming to construct a smooth interpolation between a large cut-off regime (called ultraviolet regime) and a small cut-off regime (called infrared regime).
In the large cut-off regime, fluctuations are essentially frozen and the behavior of the network is mainly fixed by the saddle point of the log-probability $S[f]$.
On the contrary, in the infrared regime, fluctuations are integrated out and look as a novel effective physics.
The \emph{Wetterich equation} describes how the effective description changes with the observation scale and leads to:
\begin{equation}
    \label{flowmassNew}
    \Lambda\frac{\dd }{\dd \Lambda}{\Gamma}_{\Lambda}^{(2)}(p, -p)
        = - \frac{1}{2} \int \frac{\dd^{d_{\text{in}}} q}{(2\pi)^{d_{\text{in}}}} \,
             \Lambda \frac{\dd {r}_\Lambda}{\dd \Lambda}(q^2) \, \Gamma_\Lambda^{(4)}(p, -p, q, -q) \, G_\Lambda^2(q^2),
\end{equation}
where $\Gamma_\Lambda^{(n)}$ is the $n$-th derivative of $\Gamma_\Lambda$ with respect to $f_{\text{cl}}$, which is defined such that:
\begin{equation}
    \Gamma_\Lambda[f_{\text{cl}}]
        := j \cdot f_{\text{cl}} - W_\Lambda[j]
            - \frac{1}{2} \, f_{\text{cl}} \cdot r_\Lambda \cdot f_{\text{cl}}, \qquad  f_{\text{cl}}(x)
        := \frac{\delta W_\Lambda}{\delta j},
\end{equation}
where
\begin{equation}
W_\Lambda[j] := \mathbb{E}[e^{-\frac{1}{2} f \cdot r_\Lambda \cdot f +j \cdot f}]\,,
\end{equation}
the dot denoting the inner-product defined by integrating over the data space. Once again, let us note that in the power field expansion of $\Gamma_\Lambda$ the weights are effective rather than bare couplings.
The \emph{regulator} $r_\Lambda$ depends on $p^2$ and is designed such that $\Gamma_{\Lambda \to \infty} \to S$ (large cut-off regime) and $\Gamma_{\Lambda \to 0} \equiv \Gamma$ (vanishing cut-off regime), $\Gamma$ being the full effective action, i.e.~the Legendre transform of the characteristic function $\mathbb{E}[e^{j \cdot f}]$.
The expectation value $K_W(p)$ being fixed by the NN, although both $\Gamma_\Lambda^{(2)}(p_1, p_2)$ and $r_\Lambda(p^2) \, \delta^{(d_{\text{in}})}(p_1 + p_2)$ can be arbitrary functions of the momentum $p^2$, their sum is constrained to be $ \Lambda^2 \, \exp \Big( \frac{p_1^2}{\Lambda^2}\Big) \, \delta^{(d_{\text{in}})}(p_1 + p_2)$ for any $\Lambda$.
Because $\Lambda\frac{\dd {r}_\Lambda}{\dd \Lambda}(q^2)$ has to select only a short window of momenta in the vicinity of the scale $\Lambda$, the smooth function $\Gamma_\Lambda^{(4)}(p, -p, q, -q)$ can be expanded in power of $q$ for $\Lambda$ small enough.
At zero order and using the Litim's regulator:
\begin{equation}
r_\Lambda(p^2) := \alpha \, (\Lambda^2 - p^2) \theta(\Lambda^2 - p^2)\,,
\end{equation}
we predict a purely scaling behavior with respect to the standard deviation $\sigma_W$ for the zero momenta function $\Gamma_{\Lambda}^{(4)}(0, 0, 0, 0) =: u_4(\Lambda) \delta(0)$:
\begin{equation}
    \label{eq:flow-u4}
    \sigma_W \, \frac{\dd u_4}{\dd \sigma_W}
        = (4 - d_{\text{in}}) u_4
    \quad \Longrightarrow \quad
    \log u_4
        = (4 - d_{\text{in}}) \log \sigma_W + \text{cst}.
\end{equation}
This equation can be verified numerically (Figure~\ref{fig:flow-u4-sw}).
A similar equation can be derived for $u_6$: $ \log u_6 = (6 -2 d_{\text{in}}) \log \sigma_W + \text{cst}$.

\section{Conclusion}
\label{sec:conclusion}

In this paper, we have reviewed the NN-QFT correspondence and described several checks.
Our main result is about the derivation of exact renormalization equations where the standard deviation $\sigma_W$ looks like a RG flow parameter, and the nice agreement between theoretical predictions and numerical experiments.

Future directions include: increasing $d_{\text{in}}$, $d_{\text{out}}$ and $N$ expansion, studying the large $d_{\text{in}}$ limit (large data), increasing number of hidden layers and extending to non-translation invariant kernels (ReLU…) using the 2PI formalism~\cite{Blaizot:2021:FunctionalRenormalizationGroup}, and finally studying the evolution of the QFT under training.

\section*{Acknowledgments}

This project has received funding from the European Union's Horizon 2020 research and innovation program under the Marie Skłodowska-Curie grant agreement No 891169.
This work is supported by the National Science Foundation under Cooperative Agreement PHY-2019786 (The NSF AI Institute for Artificial Intelligence and Fundamental Interactions, \url{http://iaifi.org/}).

\appendix

\printbibliography[heading=bibintoc]

@article{Agliari:2020:MachineLearningStatistical,
  title = {{Machine Learning and Statistical Physics: Preface}},
  shorttitle = {{Machine Learning and Statistical Physics}},
  author = {Agliari, Elena and Barra, Adriano and Sollich, Peter and Zdeborová, Lenka},
  year = {2020},
  month = nov,
  journal = {Journal of Physics A: Mathematical and Theoretical},
  volume = {53},
  number = {50},
  pages = {500401},
  publisher = {{IOP Publishing}},
  doi = {10.1088/1751-8121/abca75},
  langid = {english}
}

@article{Amit:1987:StatisticalMechanicsNeural,
  title = {{Statistical Mechanics of Neural Networks near Saturation}},
  author = {Amit, Daniel J and Gutfreund, Hanoch and Sompolinsky, H},
  year = {1987},
  month = jan,
  journal = {Annals of Physics},
  volume = {173},
  number = {1},
  pages = {30--67},
  doi = {10.1016/0003-4916(87)90092-3}
}

@article{Bahri:2020:StatisticalMechanicsDeep,
  title = {{Statistical Mechanics of Deep Learning}},
  author = {Bahri, Yasaman and Kadmon, Jonathan and Pennington, Jeffrey and Schoenholz, Sam S. and {Sohl-Dickstein}, Jascha and Ganguli, Surya},
  year = {2020},
  month = mar,
  journal = {Annual Review of Condensed Matter Physics},
  volume = {11},
  number = {1},
  pages = {501--528},
  publisher = {{Annual Reviews}},
  doi = {10.1146/annurev-conmatphys-031119-050745}
}

@article{Beny:2013:DeepLearningRenormalization,
  title = {{Deep Learning and the Renormalization Group}},
  author = {Bény, Cédric},
  year = {2013},
  month = jan,
  eprint = {1301.3124},
  archiveprefix = {arXiv},
  langid = {english}
}

@article{Beny:2018:InferringRelevantFeatures,
  title = {{Inferring Relevant Features: From QFT to PCA}},
  shorttitle = {{Inferring Relevant Features}},
  author = {Bény, Cédric},
  year = {2018},
  month = dec,
  journal = {International Journal of Quantum Information},
  volume = {16},
  number = {08},
  eprint = {1802.05756},
  pages = {1840012},
  doi = {10.1142/S0219749918400129},
  archiveprefix = {arXiv}
}

@article{Bergstra:2012:RandomSearchHyperparameter,
  title = {{Random Search for Hyper-Parameter Optimization}},
  author = {Bergstra, James and Bengio, Yoshua},
  year = {2012},
  month = feb,
  journal = {The Journal of Machine Learning Research},
  volume = {13},
  pages = {281--305},
  doi = {10.5555/2188385.2188395}
}

@article{Blaizot:2021:FunctionalRenormalizationGroup,
  title = {{Functional Renormalization Group and 2PI Effective Action Formalism}},
  author = {Blaizot, Jean-Paul and Pawlowski, Jan M. and Reinosa, Urko},
  year = {2021},
  month = feb,
  eprint = {2102.13628},
  primaryclass = {cond-mat, physics:hep-ph, physics:hep-th},
  archiveprefix = {arXiv}
}

@article{Choromanska:2015:LossSurfacesMultilayer,
  title = {{The Loss Surfaces of Multilayer Networks}},
  author = {Choromanska, Anna and Henaff, Mikael and Mathieu, Michael and Arous, Gérard Ben and LeCun, Yann},
  year = {2015},
  month = jan,
  eprint = {1412.0233},
  primaryclass = {cs},
  archiveprefix = {arXiv}
}

@article{Cranmer:2020:DiscoveringSymbolicModels,
  title = {{Discovering Symbolic Models from Deep Learning with Inductive Biases}},
  author = {Cranmer, Miles and {Sanchez-Gonzalez}, Alvaro and Battaglia, Peter and Xu, Rui and Cranmer, Kyle and Spergel, David and Ho, Shirley},
  year = {2020},
  month = nov,
  eprint = {2006.11287},
  primaryclass = {astro-ph, physics:physics, stat},
  archiveprefix = {arXiv}
}

@phdthesis{Csaji:2001:ApproximationArtificialNeural,
  title = {{Approximation with Artificial Neural Networks}},
  author = {Csáji, Balázs Csanád},
  year = {2001},
  langid = {english},
  school = {Eötvös Loránd University}
}

@article{Cybenko:1989:ApproximationSuperpositionsSigmoidal,
  title = {{Approximation by Superpositions of a Sigmoidal Function}},
  author = {Cybenko, G.},
  year = {1989},
  month = dec,
  journal = {Mathematics of Control, Signals and Systems},
  volume = {2},
  number = {4},
  pages = {303--314},
  doi = {10.1007/BF02551274},
  langid = {english}
}

@article{deMelloKoch:2020:DeepLearningRenormalization,
  title = {{Is Deep Learning a Renormalization Group Flow?}},
  author = {{de Mello Koch}, Ellen and {de Mello Koch}, Robert and Cheng, Ling},
  year = {2020},
  journal = {IEEE Access},
  volume = {8},
  eprint = {1906.05212},
  pages = {106487--106505},
  doi = {10.1109/ACCESS.2020.3000901},
  archiveprefix = {arXiv}
}

@article{Erbin:2022:NonperturbativeRenormalizationNeural,
  ids = {Erbin:2022:NonperturbativeRenormalizationNeural-1},
  title = {{Nonperturbative Renormalization for the Neural Network-QFT Correspondence}},
  author = {Erbin, Harold and Lahoche, Vincent and Samary, Dine Ousmane},
  year = {2022},
  month = mar,
  journal = {Machine Learning: Science and Technology},
  volume = {3},
  number = {1},
  eprint = {2108.01403},
  pages = {015027},
  doi = {10.1088/2632-2153/ac4f69},
  archiveprefix = {arXiv}
}

@inproceedings{Falkner:2018:BOHBRobustEfficient,
  title = {{BOHB: Robust and Efficient Hyperparameter Optimization at Scale}},
  shorttitle = {{BOHB}},
  booktitle = {{International Conference on Machine Learning}},
  author = {Falkner, Stefan and Klein, Aaron and Hutter, Frank},
  year = {2018},
  month = jul,
  pages = {1437--1446},
  publisher = {{PMLR}},
  url = {http://proceedings.mlr.press/v80/falkner18a.html},
  langid = {english}
}

@article{Gardner:1988:OptimalStorageProperties,
  title = {{Optimal Storage Properties of Neural Network Models}},
  author = {Gardner, E. and Derrida, B.},
  year = {1988},
  month = jan,
  journal = {Journal of Physics A: Mathematical and General},
  volume = {21},
  number = {1},
  pages = {271--284},
  publisher = {{IOP Publishing}},
  doi = {10.1088/0305-4470/21/1/031},
  langid = {english}
}

@article{Gardner:1988:SpaceInteractionsNeural,
  title = {{The Space of Interactions in Neural Network Models}},
  author = {Gardner, E.},
  year = {1988},
  month = jan,
  journal = {Journal of Physics A: Mathematical and General},
  volume = {21},
  number = {1},
  pages = {257--270},
  publisher = {{IOP Publishing}},
  doi = {10.1088/0305-4470/21/1/030},
  langid = {english}
}

@article{Grosvenor:2022:EdgeChaosQuantum,
  title = {{The Edge of Chaos: Quantum Field Theory and Deep Neural Networks}},
  shorttitle = {{The Edge of Chaos}},
  author = {Grosvenor, Kevin T. and Jefferson, Ro},
  year = {2022},
  month = jan,
  eprint = {2109.13247},
  primaryclass = {cond-mat, physics:hep-th, stat},
  archiveprefix = {arXiv}
}

@article{Halverson:2021:BuildingQuantumField,
  title = {{Building Quantum Field Theories Out of Neurons}},
  author = {Halverson, James},
  year = {2021},
  month = dec,
  eprint = {2112.04527},
  primaryclass = {hep-ph, physics:hep-th},
  archiveprefix = {arXiv}
}

@article{Halverson:2021:NeuralNetworksQuantum,
  title = {{Neural Networks and Quantum Field Theory}},
  author = {Halverson, James and Maiti, Anindita and Stoner, Keegan},
  year = {2021},
  month = mar,
  journal = {Machine Learning: Science and Technology},
  eprint = {2008.08601},
  doi = {10.1088/2632-2153/abeca3},
  archiveprefix = {arXiv}
}

@article{Helias:2019:StatisticalFieldTheory,
  title = {{Statistical Field Theory for Neural Networks}},
  author = {Helias, Moritz and Dahmen, David},
  year = {2019},
  month = jan,
  eprint = {1901.10416},
  archiveprefix = {arXiv}
}

@article{Henderson:2020:SystematicReportingEnergy,
  title = {{Towards the Systematic Reporting of the Energy and Carbon Footprints of Machine Learning}},
  author = {Henderson, Peter and Hu, Jieru and Romoff, Joshua and Brunskill, Emma and Jurafsky, Dan and Pineau, Joelle},
  year = {2020},
  month = jan,
  eprint = {2002.05651},
  primaryclass = {cs},
  archiveprefix = {arXiv}
}

@techreport{HighLevelExpertGrouponAI:2019:EthicsGuidelinesTrustworthy,
  title = {{Ethics Guidelines for Trustworthy AI}},
  author = {{High-Level Expert Group on AI}},
  year = {2019},
  month = apr,
  institution = {{European Commission}},
  url = {https://www.aepd.es/sites/default/files/2019-12/ai-ethics-guidelines.pdf}
}

@article{Hornik:1989:MultilayerFeedforwardNetworks,
  title = {{Multilayer Feedforward Networks Are Universal Approximators}},
  author = {Hornik, Kurt and Stinchcombe, Maxwell and White, Halbert},
  year = {1989},
  month = jan,
  journal = {Neural Networks},
  volume = {2},
  number = {5},
  pages = {359--366},
  doi = {10.1016/0893-6080(89)90020-8},
  langid = {english}
}

@article{Hornik:1991:ApproximationCapabilitiesMultilayer,
  title = {{Approximation Capabilities of Multilayer Feedforward Networks}},
  author = {Hornik, Kurt},
  year = {1991},
  month = jan,
  journal = {Neural Networks},
  volume = {4},
  number = {2},
  pages = {251--257},
  doi = {10.1016/0893-6080(91)90009-T},
  langid = {english}
}

@article{Krauth:1988:BasinsAttractionPerceptronlike,
  title = {{Basins of Attraction in a Perceptron-like Neural Network}},
  author = {Krauth, Werner and Mézard, Marc and Nadal, Jean-Pierre},
  year = {1988},
  month = aug,
  journal = {Complex Syst.},
  volume = {2},
  number = {4},
  pages = {387--408},
  doi = {10.5555/56123.56124}
}

@inproceedings{Lacoste:2019:QuantifyingCarbonEmissions,
  title = {{Quantifying the Carbon Emissions of Machine Learning}},
  booktitle = {{arXiv:1910.09700 [Cs]}},
  author = {Lacoste, Alexandre and Luccioni, Alexandra and Schmidt, Victor and Dandres, Thomas},
  year = {2019},
  month = nov,
  eprint = {1910.09700},
  primaryclass = {cs},
  archiveprefix = {arXiv}
}

@article{Leshno:1993:MultilayerFeedforwardNetworks,
  title = {{Multilayer Feedforward Networks with a Nonpolynomial Activation Function Can Approximate Any Function}},
  author = {Leshno, Moshe and Lin, Vladimir Ya. and Pinkus, Allan and Schocken, Shimon},
  year = {1993},
  month = jan,
  journal = {Neural Networks},
  volume = {6},
  number = {6},
  pages = {861--867},
  doi = {10.1016/S0893-6080(05)80131-5},
  langid = {english}
}

@article{Li:2018:HyperbandNovelBanditBased,
  title = {{Hyperband: A Novel Bandit-Based Approach to Hyperparameter Optimization}},
  shorttitle = {{Hyperband}},
  author = {Li, Lisha and Jamieson, Kevin and DeSalvo, Giulia and Rostamizadeh, Afshin and Talwalkar, Ameet},
  year = {2018},
  journal = {Journal of Machine Learning Research},
  volume = {18},
  number = {185},
  pages = {1--52},
  url = {http://jmlr.org/papers/v18/16-558.html}
}

@article{Li:2018:VisualizingLossLandscape,
  title = {{Visualizing the Loss Landscape of Neural Nets}},
  author = {Li, Hao and Xu, Zheng and Taylor, Gavin and Studer, Christoph and Goldstein, Tom},
  year = {2018},
  month = nov,
  eprint = {1712.09913},
  primaryclass = {cs, stat},
  archiveprefix = {arXiv}
}

@article{Lin:2017:WhyDoesDeep,
  title = {{Why Does Deep and Cheap Learning Work so Well?}},
  author = {Lin, Henry W. and Tegmark, Max and Rolnick, David},
  year = {2017},
  journal = {Journal of Statistical Physics},
  volume = {168},
  number = {6},
  eprint = {1608.08225},
  pages = {1223--1247},
  doi = {10.1007/s10955-017-1836-5},
  archiveprefix = {arXiv},
  langid = {english}
}

@incollection{Lu:2017:ExpressivePowerNeural,
  title = {{The Expressive Power of Neural Networks: A View from the Width}},
  shorttitle = {{The Expressive Power of Neural Networks}},
  booktitle = {{Advances in Neural Information Processing Systems 30}},
  author = {Lu, Zhou and Pu, Hongming and Wang, Feicheng and Hu, Zhiqiang and Wang, Liwei},
  editor = {Guyon, I. and Luxburg, U. V. and Bengio, S. and Wallach, H. and Fergus, R. and Vishwanathan, S. and Garnett, R.},
  year = {2017},
  eprint = {1709.02540},
  pages = {6231--6239},
  publisher = {{Curran Associates, Inc.}},
  url = {http://papers.nips.cc/paper/7203-the-expressive-power-of-neural-networks-a-view-from-the-width.pdf},
  archiveprefix = {arXiv}
}

@article{Maiti:2021:SymmetryviaDualityInvariantNeural,
  title = {{Symmetry-via-Duality: Invariant Neural Network Densities from Parameter-Space Correlators}},
  shorttitle = {{Symmetry-via-Duality}},
  author = {Maiti, Anindita and Stoner, Keegan and Halverson, James},
  year = {2021},
  month = jun,
  eprint = {2106.00694},
  archiveprefix = {arXiv}
}

@article{Mehta:2014:ExactMappingVariational,
  title = {{An Exact Mapping between the Variational Renormalization Group and Deep Learning}},
  author = {Mehta, Pankaj and Schwab, David J.},
  year = {2014},
  month = oct,
  eprint = {1410.3831},
  archiveprefix = {arXiv}
}

@article{Mezard:1989:LearningFeedforwardLayered,
  title = {{Learning in Feedforward Layered Networks: The Tiling Algorithm}},
  shorttitle = {{Learning in Feedforward Layered Networks}},
  author = {Mézard, M. and Nadal, Jean-P.},
  year = {1989},
  journal = {Journal of Physics A: Mathematical and General},
  volume = {22},
  number = {12},
  pages = {2191},
  doi = {10.1088/0305-4470/22/12/019},
  langid = {english}
}

@book{Neal:1996:BayesianLearningNeural,
  title = {{Bayesian Learning for Neural Networks}},
  author = {Neal, Radford M.},
  year = {1996},
  series = {Lecture {{Notes}} in {{Statistics}}},
  publisher = {{Springer-Verlag}},
  doi = {10.1007/978-1-4612-0745-0},
  langid = {english}
}

@article{Pinkus:1999:ApproximationTheoryMLP,
  title = {{Approximation Theory of the MLP Model in Neural Networks}},
  author = {Pinkus, Allan},
  year = {1999},
  month = jan,
  journal = {Acta Numerica},
  volume = {8},
  pages = {143--195},
  publisher = {{Cambridge University Press}},
  doi = {10.1017/S0962492900002919},
  langid = {english}
}

@article{Raghu:2017:ExpressivePowerDeep,
  title = {{On the Expressive Power of Deep Neural Networks}},
  author = {Raghu, Maithra and Poole, Ben and Kleinberg, Jon and Ganguli, Surya and {Sohl-Dickstein}, Jascha},
  year = {2017},
  month = jun,
  eprint = {1606.05336},
  primaryclass = {cs, stat},
  archiveprefix = {arXiv}
}

@article{Raghu:2020:SurveyDeepLearning,
  title = {{A Survey of Deep Learning for Scientific Discovery}},
  author = {Raghu, Maithra and Schmidt, Eric},
  year = {2020},
  month = mar,
  eprint = {2003.11755},
  primaryclass = {cs, stat},
  archiveprefix = {arXiv}
}

@article{Roscher:2020:ExplainableMachineLearning,
  title = {{Explainable Machine Learning for Scientific Insights and Discoveries}},
  author = {Roscher, Ribana and Bohn, Bastian and Duarte, Marco F. and Garcke, Jochen},
  year = {2020},
  journal = {IEEE Access},
  volume = {8},
  pages = {42200--42216},
  doi = {10.1109/ACCESS.2020.2976199}
}

@book{Saitta:2011:PhaseTransitionsMachine,
  title = {Phase Transitions in Machine Learning},
  author = {Saitta, Lorenza and Giordana, Attilio and Cornuéjols, Antoine},
  year = {2011},
  month = jun,
  publisher = {{Cambridge University Press}},
  langid = {Anglais}
}

@article{Schoenholz:2017:CorrespondenceRandomNeural,
  title = {{A Correspondence Between Random Neural Networks and Statistical Field Theory}},
  author = {Schoenholz, Samuel S. and Pennington, Jeffrey and {Sohl-Dickstein}, Jascha},
  year = {2017},
  month = oct,
  eprint = {1710.06570},
  archiveprefix = {arXiv}
}

@misc{Sharir:2020:CostTrainingNLP,
  title = {{The Cost of Training NLP Models: A Concise Overview}},
  shorttitle = {{The Cost of Training NLP Models}},
  author = {Sharir, Or and Peleg, Barak and Shoham, Yoav},
  year = {2020},
  month = apr,
  number = {arXiv:2004.08900},
  eprint = {2004.08900},
  primaryclass = {cs},
  publisher = {{arXiv}},
  doi = {10.48550/arXiv.2004.08900},
  archiveprefix = {arXiv}
}

@inproceedings{Snoek:2012:PracticalBayesianOptimization,
  title = {{Practical Bayesian Optimization of Machine Learning Algorithms}},
  booktitle = {{Proceedings of the 25th International Conference on Neural Information Processing Systems - Volume 2}},
  author = {Snoek, Jasper and Larochelle, Hugo and Adams, Ryan P.},
  year = {2012},
  month = dec,
  series = {{{NIPS}}'12},
  eprint = {1206.2944},
  pages = {2951--2959},
  publisher = {{Curran Associates Inc.}},
  doi = {10.5555/2999325.2999464},
  archiveprefix = {arXiv}
}

@misc{Strubell:2019:EnergyPolicyConsiderations,
  title = {{Energy and Policy Considerations for Deep Learning in NLP}},
  author = {Strubell, Emma and Ganesh, Ananya and McCallum, Andrew},
  year = {2019},
  month = jun,
  number = {arXiv:1906.02243},
  eprint = {1906.02243},
  primaryclass = {cs},
  publisher = {{arXiv}},
  doi = {10.48550/arXiv.1906.02243},
  archiveprefix = {arXiv}
}

@article{Weld:2018:ChallengeCraftingIntelligible,
  title = {{The Challenge of Crafting Intelligible Intelligence}},
  author = {Weld, Daniel S. and Bansal, Gagan},
  year = {2018},
  month = oct,
  eprint = {1803.04263},
  primaryclass = {cs},
  archiveprefix = {arXiv}
}

@article{Yang:2021:TensorProgramsWide,
  title = {{Tensor Programs I: Wide Feedforward or Recurrent Neural Networks of Any Architecture Are Gaussian Processes}},
  shorttitle = {{Tensor Programs I}},
  author = {Yang, Greg},
  year = {2021},
  month = may,
  eprint = {1910.12478},
  primaryclass = {cond-mat, physics:math-ph},
  archiveprefix = {arXiv}
}

@article{Zdeborova:2020:UnderstandingDeepLearning,
  title = {{Understanding Deep Learning Is Also a Job for Physicists}},
  author = {Zdeborová, Lenka},
  year = {2020},
  month = may,
  journal = {Nature Physics},
  pages = {1--3},
  publisher = {{Nature Publishing Group}},
  doi = {10.1038/s41567-020-0929-2},
  copyright = {2020 Springer Nature Limited},
  langid = {english}
}

@article{Zhang:2021:SurveyNeuralNetwork,
  title = {{A Survey on Neural Network Interpretability}},
  author = {Zhang, Yu and Tiňo, Peter and Leonardis, Aleš and Tang, Ke},
  year = {2021},
  month = oct,
  journal = {IEEE Transactions on Emerging Topics in Computational Intelligence},
  volume = {5},
  number = {5},
  eprint = {2012.14261},
  primaryclass = {cs},
  pages = {726--742},
  doi = {10.1109/TETCI.2021.3100641},
  archiveprefix = {arXiv}
}

\end{document}